\documentclass[aps,reprint,amsmath,amssymb,superscriptaddress,nobibnotes,showpacs,showkeys,twocolumn]{revtex4-1}

\usepackage{graphicx}
\usepackage{dcolumn}
\usepackage{bm}
\usepackage{amsmath}
\usepackage{amssymb,amsthm}
\usepackage{bbm}
\usepackage{epsfig,color}
\usepackage[sans]{dsfont}

\usepackage{graphicx}
\graphicspath{ {images/} }

\definecolor{nblue}{rgb}{0.2,0.2,0.7}
\definecolor{ngreen}{rgb}{0.2,0.6,0.2}
\definecolor{nred}{rgb}{0.7,0.2,0.2}
\definecolor{nblack}{rgb}{0,0,0} 

\newcommand{\tr}{\text{tr}}

\renewcommand{\H}{\mathcal{H}}

  \theoremstyle{definition}
  
  \theoremstyle{plain}
  
\theoremstyle{plain}
  
\theoremstyle{plain}

  \theoremstyle{plain}
  
  \theoremstyle{plain}

  \providecommand{\conjecturename}{Conjecture}
  \providecommand{\definitionname}{Definition}
  \providecommand{\lemmaname}{Lemma}
\providecommand{\corollaryname}{Corollary}
\providecommand{\theoremname}{Theorem}
\providecommand{\propositionname}{Proposition}

\def\ew{\widetilde{W}}

\def\i{\mathrm{id}}

\def\H{\mathcal{H}}

\def\tr{\mbox{tr}}

\def\bea{\begin{eqnarray}}
\def\eea{\end{eqnarray}}

\newtheorem{Example}{Example}

\begin{document}



\title{Entanglement Witness $2.0$: Compressed/Mirrored Entanglement Witnesses}

\author{Joonwoo Bae}
\affiliation{School of Electrical Engineering, Korea Advanced Institute of Science and Technology (KAIST), 291
Daehak-ro, Yuseong-gu, Daejeon 34141, Republic of Korea}
\email{joonwoo.bae@kaist.ac.kr}

\author{Dariusz Chru\'sci\'nski}
\affiliation{Institute of Physics, Faculty of Physics, Astronomy, and Informatics, Nicolaus Copernicus University, Grudziadzka 5, 87-100 Torun, Poland}
\email{darch@fizyka.umk.pl}

\author{Beatrix C. Hiesmayr}
\affiliation{University of Vienna, Faculty of Physics, Boltzmanngasse 5, 1090 Vienna, Austria}
\email{Beatrix.Hiesmayr@univie.ac.at}


\begin{abstract}
Entanglement detection, that signifies the task of distinguishing entangled states from separable states, can be generally performed by realizing entanglement witnesses via local measurements on a single-copy level and classical communication, and are known to be experimenter friendly. We introduce a framework of constructing {\it mirrored entanglement witnesses} by showing that an experimental observable is twice as effective since it generally provides bounds from above and below for separable states. Differently stated, a pair of witnesses, {\it mirrored witnesses}, exist for the characterization of the bounds, which are two faces of one observable. We show how to generally construct those witnesses and provide examples for bipartite and multipartite systems. We also show that both mirrored witnesses can be improved by introducing nonlinearities, by which a larger set of entangled states can be certified.


\end{abstract}

\maketitle

\section{Introduction}

Revealing properties of an unknown quantum state of a physical system is of fundamental importance in quantum information theory and its application. By quantum state tomography, the full knowledge of a given physical system can be determined after series of quantum measurements, which is, however, generally an expensive and cost-inefficient procedure demanding experimental resources. Moreover, one is often only interested in the question of whether a multipartite state is entangled or not and what type of entanglement is present, for instance bipartite or genuinely multipartite entanglement. Entanglement witnesses (EWs), particular observables, offer the solution with currently feasible technologies. EWs typically require less experimental setups to unambiguously verify the existence of entanglement and even additional information on the type of entanglement may be revealed. This concept relies on the fact that the set of separable states are convex, i.e. EWs correspond to hyperplanes separating some entangled states from the set of separable states in the Hilbert-Schmidt space~\cite{ref:terhal}. In high-dimensional and multi-partite systems also the entanglement structure itself becomes of interest which has a nested convex structure~\cite{ref:review1,ref:review2,ref:review3}. In the recent years, there has been much progress in the theory and experimental realization of EWs for multipartite entangled states, see e.g., \cite{ref:new1, ref:new2, ref:new3}.

\begin{figure}[t]
\includegraphics[width=3.5in,keepaspectratio]{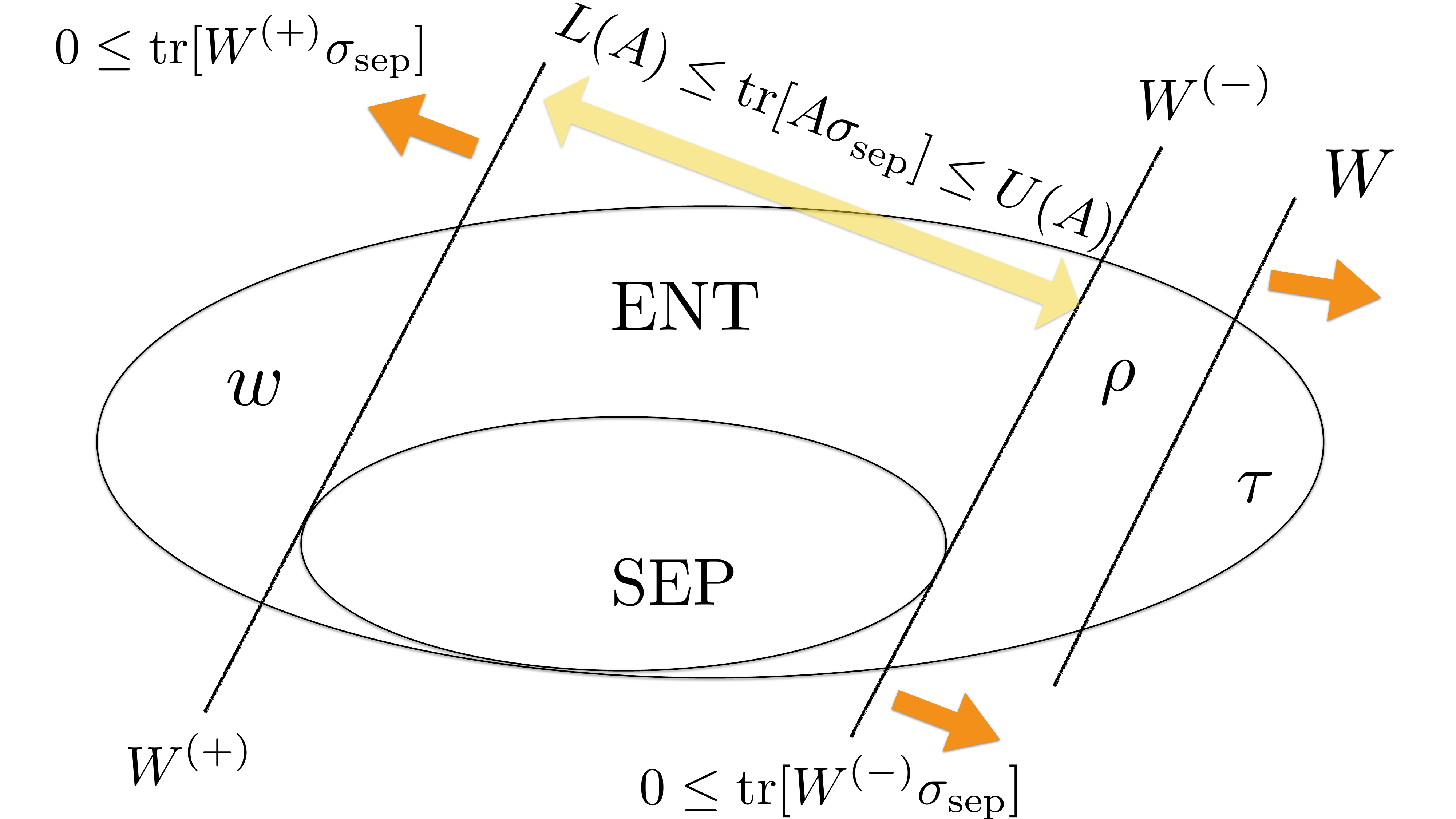}
\caption{ The framework of mirrored EWs proves that a single observable $A$ has both lower and upper bounds for separable states, where the bounds are characterized by mirrored EWs $W^{(\pm)}$. An experimental realization of a single observable $A$ can detect those entangled states, e.g. $\rho$, $\tau$, $w$, etc., which are detected by both of the mirrored EWs. }
\label{fig:pic}
\end{figure}

EWs exhibit numerous advantages. They are observables on a single-copy level and are factored into local observables, namely no entanglement resources are needed to realize EWs. These  properties makes them suited for experimental investigations. A major drawback is that \textit{a priori} information about entangled states to be detected is needed to be at hand in advance. This can be explained in Fig.~\ref{fig:pic} as follows. Suppose that an entangled state $\tau$ is detected by a witness $W$, which however does not detect states $\rho$ and $w$. The witness $W$ can be optimized to construct a finer one $W^{(-)}$, that can detect a state $\rho$ but not yet a state $w$ at the other side. Therefore, in order to detect a state $w$, {\it a priori} information about the state is needed such that a relevant witness $W^{(+)}$ can be constructed. Or, to detect a state $\rho$, one may also need a proper optimization technique to lift $W$ to $W^{(-)}$ \cite{Optimal}.

In this work, we establish the framework of {\it mirrored EWs} that can construct a pair of EWs from an EW, and show that both of the EWs can perform detecting entangled states by realizing a single observable. In other words, an EW $W^{(+)}$ has its `twin' one denoted by $W^{(-)}$, called a mirrored witness, such that the pair of EWs can be realized by a single observable. In Fig.~\ref{fig:pic}, this can be illustrated that EWs $W^{(+)}$ and $W^{(-)}$ serve the characterization of lower and upper bounds of an observable $A$ for separable states. Then, once an observable $A$ is realized, entangled states are detected from violations of the lower or the upper bounds, which are equivalent to entanglement detection by $W^{(+)}$ or $W^{(-)}$, respectively. Note that both a state $\rho$, which is not detected by $W^{(+)}$, and a state $w$, not by $W^{(-)}$, can be detected by realizing a single observable $A$. Hence, introducing mirrored EWs, one can {\em double} the capability of detecting entangled states. We show that mirrored EWs can be constructed from upper and lower bounds of an observable for separable states also present how mirrored EWs are related to the structural physical approximation. Examples are presented for the illustration of mirrored EWs. Finally, we investigate the construction of nonlinear EWs from mirrored witnesses and show that mirrored EWs can be improved by both subtraction and addition of nonlinear polynomials.  

Let us review basic properties of entanglement witnesses. Throughout, let $W$ denote an EW such that
\bea
\forall \sigma \in S_{\mathrm{sep}}, ~\tr[W\sigma ]  \geq 0, ~\mathrm{and} ~~\exists \rho \notin S_{\mathrm{sep}},~  \tr[W \rho] < 0,~~~~ \label{eq:lb}
\eea
where $S_{\mathrm{sep}}$ is the set of separable states. Assuming bipartite systems, an EW can be factorized into local measurements, i.e., $W = \sum_{i} c_i ~M_{i}^{(A)} \otimes N_{i}^{(B)}$ for some constants $\{ c_i\} $ and positive operator-valued measure (POVM) $\{ M_{i}^{(A)} \}$ and $\{ N_{i}^{(B)} \}$, that give descriptions of measurement devices. Note that an EW can be generally factorized into local measurements \cite{ref:localEW}. We can restrict the consideration to normalized EW without loss of generality, i.e., $\tr W=1$, on a Hilbert space $\H = \H_1 \otimes \H_2$ where $d_i = \rm{dim} \H_i$ and $D =d_1 d_2$. 

Recall, that if $W$ is an EW then geometrically it gives rise to a hyperplane $\Sigma$ in $D$-dimensional real space defined by $\Sigma=\{ A=A^\dagger\, |\, {\rm tr}(WA)=0\}$. A convex set of separable states is located only on one side of $\Sigma$ (cf. Fig. 1). Denote by $D(W)$ a set of states detected by $W$. One calls \cite{Optimal} EW $W'$ finer than $W$ if $D(W) \subset D(W')$. Finally, $W$ is optimal if there is no witness finer than $W$. Clearly, when $W$ is optimal the corresponding hyperplane $\Sigma$ touches a set of separable states (in Fig.~\ref{fig:pic} the EW $W^{(-)}$ is \textit{finer} than $W$ since it detects all entangled states of $W$ and more).  It is clear that any EW $W$ can be shifted by subtracting a positive operator such that the corresponding $\Sigma$ touches the border of separable states. An EW $W$ is called \textit{decomposable} if it has a form $W = A + B^\Gamma$ for $A,B \geq 0$, where $B^\Gamma$ denotes the partial transposition on the operator $B$. An EW which is not of this form is called non-decomposable and can detect bound entangled states. Bound entanglement has been detected experimentally for multipartite systems, such as four-qubit cases with polarization degrees of freedom~\cite{BE1, BE2}. Bipartite bound entanglement is the strongest form of bound entanglement and in 2013 it has been experimentally demonstrated by photons entangled in their orbital angular momentum degrees of freedom~\cite{HL2013}.


Using the well-known duality between linear maps $\Phi: \mathcal{B}(\mathcal{H}_1) \to \mathcal{B}(\mathcal{H}_2)$ and operators in $\mathcal{H}_1 \otimes \mathcal{H}_2$ defined by Choi-Jamio{\l}kowski isomorhism
\begin{equation}\label{CJ}
  W = ({\rm id} \otimes \Phi)(P^+_1)  ,
\end{equation}
where $P^+_1$ denotes the maximally entangled state in $\mathcal{H}_1 \otimes \mathcal{H}_1$, and `${\rm id}$' is an identity map in  $\mathcal{B}(\mathcal{H}_1)$, one may immediately translate all properties of EWs into the corresponding properties of maps and {\it vice versa}. In Eq.~(\ref{CJ}), $W$ is an EW if and only if $\Phi$ is a positive but not completely positive map.


\section{Mirrored entanglement witnesses}

Consider a Hermitian operator $A\geq 0$ in ${\cal H}_1 \otimes {\cal H}_2$ such that $\tr A=1$, that can be interpreted as a state of a composite system. Let $\lambda_{\max} (A)$ and $\lambda_{\min} (A)$ denote the maximal and minimal eigenvalues of $A$, respectively. Moreover, we introduce  the upper ($U$) and lower ($L$) borders
\bea
 U(A) = \max_{\sigma \in S_{\rm sep}} \,  {\rm tr} [ A \sigma ]~\mathrm{and}~  L(A) = \min_{\sigma \in S_{\rm sep}} \,  \tr  [A \sigma  ]\; , \nonumber
\eea
 such that for any separable state $\sigma \in S_{\mathrm{sep}}$ follows
\bea
L(A) \leq \tr[A  \sigma ] \leq  U(A). \label{eq:main}
\eea
Now, if $\lambda_{\min}(A)< L(A)$ and $\lambda_{\max}(A) > U(A)$, both lower and upper bounds may be used to detect entangled states: if one finds that for a state $\rho$
\bea  \tr[ A \rho] \notin [L(A) ,U(A)] ,\label{eq:maind}
\eea
then the state is entangled. We call the interval $[L(A) ,U(A)]$ in Eq.~(\ref{eq:main}) the {\em separability window} of an observable $A$.


A separability window of an observable may be found in terms of {\em separability eigenvalue equations} as follows~\cite{Vogel}:
\bea
  A_1 |x_1\rangle = \lambda |x_1\rangle , \ \ \   A_2 |x_2\rangle = \lambda |x_2\rangle , \label{vogel}
\eea
where $A_1 = {\rm tr}_2 [\mathrm{I}_1 \otimes |x_2\rangle \langle x_2| \cdot A] \in \mathcal{B}(\mathcal{H}_1)$ and $A_2 = {\rm tr}_1 [|x_1\rangle \langle x_1| \otimes \mathrm{I}_2 \cdot  A] \in \mathcal{B}(\mathcal{H}_2)$.  Let ${\lambda}_{1}$ and ${\lambda}_{2}$ denote minimal and maximal eigenvalues, respectively, in Eq.~(\ref{vogel}). Then, the separability window of an observable $A$ can be found as, $[L(A) ,U(A)] = [ {\lambda}_{1},{\lambda}_{2}]$.


As the scheme of detecting entangled states with a separability window in Eq. (\ref{eq:maind}) is linear with respect to quantum states, one may anticipate that both of the bounds may be identified by EWs. In fact, we construct a pair of EWs that can equivalently characterize quantum states within a nontrivial separability window $[L(A),U(A)]$ for an observable $A$, as follow
\begin{eqnarray}
   \label{M} W^{(-)} &=&  \frac{1}{n_-} \Big[ U(A) \; \mathrm{I}_1 \otimes \mathrm{I}_2- A \Big]\,,  \label{eq:w-a} \\
   \label{P} W^{(+)} &=& \frac{1}{n_+} \Big[ A - L(A) \; \mathrm{I}_1 \otimes \mathrm{I}_2 \Big]\,,  \label{eq:w+a}
\end{eqnarray}
where the normalization factors read
\bea
n_- =  D\; U(A)-1 \ \ , \ \   n_+ = 1- D\;L(A) , \label{eq:const}
\eea
and $D=d_1\cdot d_2$. One can immediately prove that $\tr [W^{(\pm)} \sigma  ] \geq 0 $
for all $\sigma \in S_{\mathrm{sep}}$ by referring to Eq. (\ref{eq:main}), i.e., for a state $\rho$,
\bea
\tr[A\rho] \leq U(A) & \iff & \tr[W^{(-)} \rho ] \geq 0,~\mathrm{and} \nonumber \\
\tr[A\rho] \geq L(A) & \iff & \tr[W^{(+)} \rho ] \geq 0. \label{eq:equi}
\eea
We call a pair of EWs $(W^{(-)},W^{(+)})$ {\em mirrored entanglement witnesses}: they are constructed by an observable that has a nontrivial separability window. That is, entanglement detection with both mirrored EWs $W^{(\pm)}$ can be equivalently performed by realizing a single observable $A$, as it is shown in Eq.~(\ref{eq:maind}). \\

{\bf Remark.} Entangled states detected by a pair of EWs that are mirrored to each other can be realized by a single observable, that is non-negative and of unit-trace, i.e., thus can be interpreted as a quantum state.\\

\begin{figure}[t]
\includegraphics[width=2.5in,keepaspectratio]{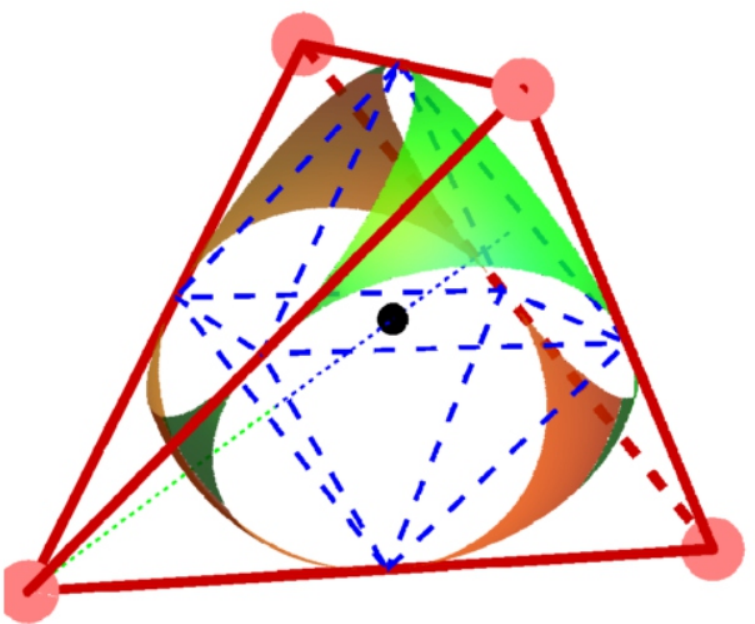}
\caption{(Color online) \textit{Visualization}  of the `{\bf Two-Qubit EW}': This magic simplex~\cite{simplex1,simplex2,simplex3} represents the Hilbert space of all bipartite qubit states which partial traces correspond to the maximally mixed states, so called \textit{locally maximally mixed states}. They can be written in the form $\rho=\frac{1}{4} (\mathrm{I\otimes I}+c_i \sigma_i\otimes\sigma_i)$ with $c_i\in \mathds{R}$. All points $\{c_1,c_2,c_3\}$ inside the (red) tetrahedron represent states (satisfy the positivity condition). All states inside the (blue) double pyramid are separable states. The (pink) dotes correspond to the four Bell states and the dotted line represents one isotropic state $\rho_\alpha$ (SEP (blue): $\alpha\geq \frac{2}{3}$; ENT (green) $\alpha<\frac{2}{3}$). The (light and dark) green surface represents $\tr[W^{(+)}\rho]=0$ and the (light and dark) orange surface $\tr[W^{(-)}\rho]=0$. In this case, the single observable that contains mirrored EWs $W^{(\pm)}$ covers the full space of all locally maximally mixed states. However, note that these witnesses are not optimal.}
\label{fig:simplex}
\end{figure}

It follows that for mirrored EWs, the corresponding hyperplanes in the Hilbert-Schmidt space, denoted by $\Sigma^{(\pm)}$, touch a set of separable states (cf. Fig.~\ref{fig:pic}), see also Eq.~(\ref{eq:main}). This gives raise to a universal relation for mirrored EWs
\bea
n_+W^{(+)} + n_-W^{(-)} = \Delta(A) \, \mathrm{I}_1 \otimes \mathrm{I}_2 ,  \label{UNI}
\eea
with $\Delta(A) =  U(A)-L(A)$ from the separability window of an observable $A$.

From the relations in Eqs.~(\ref{eq:w-a}), (\ref{eq:w+a}) and (\ref{UNI}), mirrored EWs can be directly constructed from each other, without referring to an observable. It is clear that an EW $W^{(+)}$ has its lower bound $L(W^{(+)})=0$ for separable states. It follows that its mirrored EW $W^{(-)}$ is given as,
\bea
W^{(-)} = \frac{1}{m_-}\Big[U(W^{(+)}) \, \mathrm{I}_1 \otimes \mathrm{I}_2 - W^{(+)} \Big], \nonumber
\eea
where $m_-$ stands for normalization factor. Equivalently, one finds
\bea
W^{(+)} = \frac{1}{m_+}\Big[U(W^{(-)}) \, \mathrm{I}_1 \otimes \mathrm{I}_2 - W^{(-)} \Big], \nonumber
\eea
and hence
\bea
(m_- +1)W^{(-)} + (m_+ +1)W^{(+)} = m \, \mathrm{I}_1 \otimes \mathrm{I}_2 , \nonumber
\eea
with $  m =  (m_+ +m_-+2)/D = U(W^{(-)})+U(W^{(+)})$.


\section{Mirrored structural physical approximations}

It turns out that our approach  is directly related to the structural physical approximation (SPA)~\cite{ref:spareview,ref:bae} of EWs. Recall, that for  an EW $W$ its SPA can be written as a non-negative observable
\begin{equation}\label{p}
P= (1-p) W +  p \frac{\mathrm{I_1 \otimes I_2 }}{D} 
\end{equation}
where  $p \in (0,1)$ is a minimal number such that that $P \geq 0$. Let us call a positive operator $P$ a positive SPA (p-SPA) in the sense that an EW $W$ is admixed with a positive fraction $1-p>0$.  
In a similar vein, we introduce the negative SPA (n-SPA) with a negative fraction of an EW $W$ as follows
\begin{equation}\label{q}
Q =  (1-q) W +  q \frac{\mathrm{I_1\otimes I_2}}{D} 
\end{equation}
with maximal $q > 1$ such that $Q \geq 0$. Clearly, both $P$ and $Q$ are density operators in ${\cal H}_1 \otimes {\cal H}_2$ belonging to the boundary of space of states. We call a pair $(P,Q)$ mirrored SPAs to an EW $W$. \\

{\bf Proposition}. Let $(W^{(-)},W^{(+)} )$ be mirrored EWs. Then, p-SPA to $W^{(+)}$ coincides with n-SPA to $W^{(-)}$. \\

{\bf Proof.} Let us begin with p-SPA to $W^{(+)}$ and n-SPA to $W^{(-)}$,
\bea
\ew = (1-p_{\pm}) W^{(\pm)} + p_{\pm} \frac{\mathrm{I_1\otimes I_2}}{D}, \nonumber
\eea
with $ p_{+} < 1 < p_{-}$ and $D = d_1 d_2$. Then, it holds that for any $\sigma \in S_{\mathrm{sep}}$
\bea
p_{+}/D  ~\leq ~   {\rm tr}[\ew \sigma]  ~\leq ~  p_{-}/D\;. \label{eq:prop1}
\eea
which shows $L (\ew)= p_{+}/D$ and $U(\ew) = p_{-}/D$. $\Box$\\

The result in Eq. (\ref{eq:prop1}) can be compared to Eq. (\ref{eq:main}) by replacing an observable $A$ with the resulting non-negative operator $\widetilde{W}$ after SPA. Note also that the universal relation (\ref{UNI}) reads
\bea
(p_{-}-1)  W^{(-)} + (1-p_{+})  W^{(+)} = (p_{-} - p_{+}) \frac{\mathrm{I_1\otimes I_2}}{D} .  \nonumber
\eea
Note also that mirrored EWs can be constructed by replacing $A$ in Eqs. (\ref{eq:w-a}) and  (\ref{eq:w+a}) with the operator $\widetilde{W}$ after p- (n-) SPA of $W^{(+)}$ ($W^{(-)}$).

Then, if $\ew$ belongs to the boundary of $S(\mathcal{H})$, i.e. $\lambda_{\rm min}(\ew)=0$, then  $ W^{(-)}$ detects an entanglement of $\rho = n(\mathrm{I}_1 \otimes \mathrm{I}_2 - \Pi)$, where $\Pi$ is a projector onto a support of $\ew$ and $n$ is a normalization factor. Indeed, one has ${\rm tr}( W^{(-)} \rho) = - L(\ew) n/n_- < 0$. On the other hand, for ${\rm tr}\ew^2 > U(\ew)$, we have ${\rm tr}(W^{(-)} \ew) < 0$, and hence $W^{(-)}$ detects entanglement of $\ew$.

The framework may be equivalently applied to completely positive (CP) maps. To simplify the discussion let us restrict to trace preserving maps, that is, a CP map $\widetilde{\Phi}$ corresponds to a quantum channel. Recall that a linear map $\Phi$ is positive if ${\rm tr}(P\Phi(Q))\geq 0 $ for any pair of rank-1 projectors $P$ and $Q$. A bipartite state $\rho$ is separable if $[{\rm id} \otimes \Phi](\rho) \geq 0$ for all positive maps $\Phi$ (where `id' stands for the identity map, that is, ${\rm id}(X)=X$).  Now, using Choi-Jamio{\l}kowski isomorphism, a map $\widetilde{\Phi}$ is CP if the following bipartite operator $\widetilde{W} = \sum_{i,j} |i\rangle \langle j| \otimes \widetilde{\Phi}( |i\rangle \langle j|)$ is positive. Noting that

$$   {\rm tr}(\widetilde{W} \, P^T \otimes Q) = {\rm tr}[P \widetilde{\Phi}(Q)] , $$
let us introduce the following bounds
\begin{equation}
  L(\widetilde{\Phi}) = \min {\rm tr}[P \widetilde{\Phi}(Q)]\ , \ \ \  U(\widetilde{\Phi}) = \max {\rm tr}[P \widetilde{\Phi}(Q)] , \nonumber
\end{equation}
where minimization (maximization) runs over rank-$1$ projectors $P$ and $Q$. Then, we construct positive maps as follows
\begin{eqnarray}
   \Phi^{(-)} &=&  \frac{1}{m_-} \Big[ d_2 U(\widetilde{\Phi}) \, \Phi_* - \widetilde{\Phi} \Big] ,   \nonumber\\
   \Phi^{(+)} &=& \frac{1}{m_+} \Big[ \widetilde{\Phi} - d_2 L(\widetilde{\Phi}) \, \Phi_* \Big],  \nonumber
\end{eqnarray}
with the depolarization channel $\Phi_*(\rho) = \mathrm{I}_2  {\rm tr}\rho /{d_2}$,  $m_- =  d_2 U(\widetilde{\Phi}) -1$, and $m_+ = 1- d_2 L(\widetilde{\Phi})$.  For any separable state $\sigma \in S_{\mathrm{sep}}$, one has
\bea
 L(\widetilde{\Phi}) \sigma_1 \otimes  \mathrm{I}_2 \leq   ({\rm id} \otimes \widetilde{\Phi})(\sigma) \leq U(\widetilde{\Phi}) \sigma_1 \otimes  \mathrm{I}_2 , \nonumber
\eea
where $\sigma_1 = {\rm tr}_2 \sigma$ denotes a reduced state. Similarly
\bea
 L(\widetilde{\Phi}) \mathrm {I}_1 \otimes \sigma_2 \leq   (\widetilde{\Phi} \otimes {\rm id})(\sigma) \leq U(\widetilde{\Phi}) \mathrm {I}_1 \otimes \sigma_2 ,\nonumber
\eea
with $\sigma_2 = {\rm tr}_1 \sigma$.

An example of such construction has been considered in Ref.~\cite{REMIK}. We recall that a set of product vectors $|\psi_j\rangle = |\alpha_j\rangle \otimes |\beta_j\rangle$ ($j=1,\ldots,K$) defines an unextendible product basis (UPB) in $\mathcal{H}_1 \otimes \mathcal{H}_2$ if there is no product vector orthogonal to all $|\psi_j \rangle$~\cite{UPB}. Let  us choose the projector onto UPB, $\Pi_{\rm UPB}= |\psi_1\rangle\langle \psi_1| + \ldots + |\psi_K\rangle\langle \psi_K|$. Then, an EW in the following
\begin{equation}
W = (\Pi_{\rm UPB} - L(\Pi_{\rm UPB}) \mathrm{I}_1 \otimes \mathrm{I}_2)/[K- L(\Pi_{\rm UPB}) D] \label{eq:upbew}
\end{equation}
shares precisely the same structure with $W^{(+)}$. The witness detects a bound entangled state $\rho= (\mathrm{I}_1 \otimes \mathrm{I}_2 - \Pi_{\rm UPB})/(D-K)$.


The presented framework can be immediately generalized for the multipartite case. Clearly, multipartite case is much more subtle since there are several level of separability \cite{ref:review2,ref:review3}. The simplest notion is the full separability corresponding to pure product vectors $|\Psi\rangle = |\psi_1\rangle \otimes \ldots \otimes |\psi_N\rangle \in \mathcal{H}= \mathcal{H}_1 \otimes \ldots \otimes \mathcal{H}_N$. Now for any $\ew \geq 0$ in $\mathcal{H}$ one defines $U(\ew) = \max_{|\Psi\rangle} \langle \Psi|\ew|\Psi\rangle$ and $L(\ew)= \min_{|\Psi\rangle} \langle \Psi|\ew|\Psi\rangle$, where it suffices to consider the optimization over product states. Finally one defines a pair of EWs via (\ref{M}) and (\ref{P}), generalized to multipartite systems (see also Ref. \cite{ref:sperlingm}). For example, the EW  in Eq. (\ref{eq:upbew}) can be immediately generalized as follows ~\cite{REMIK,Vogel}. Let $|\Psi_j\rangle \in \mathcal{H}$ ($j=1,\ldots,K$) denote UPB and $\Pi_{\rm UPB}= |\Psi_1\rangle\langle \Psi_1| + \ldots + |\Psi_K\rangle\langle \Psi_K|$, from which one can construct a witness $W^{(+)}  = (\Pi_{\rm UPB} - L(\Pi_{\rm UPB}) \mathrm{I})/[K- L(\Pi_{\rm UPB}) D]$, with $\mathrm{I} = \mathrm{I}_1 \otimes \ldots \mathrm{I}_N$ and $D=d_1\ldots d_N$.

\section{Examples and discussions}\label{sec:sum}

In this section, we illustrate the systematic construction of mirrored EWs with examples and then give a summary and outlook.

\begin{Example}[Two-Qubit EWs] {\em Consider the following EWs for two-qubit states
$$  W^{(+)} = \frac 18 \left( \begin{array}{cc|cc} 1 & 0 & 0 & -4 \\ 0 & 3 & 0 & 0 \\ \hline 0 & 0 & 3 & 0 \\ -4 & 0 & 0 & 1 \end{array} \right)\ , \
W^{(-)} = \frac 18 \left( \begin{array}{cc|cc} 3 & 0 & 0 & 4 \\ 0 & 1 & 0 & 0 \\ \hline 0 & 0 & 1 & 0 \\ 4 & 0 & 0 & 3 \end{array} \right), $$
which are written as block matrices in the computational basis for qubit states, e.g. $\mathrm{Diag}[1,1,1,1] = \sum_{i,j=0,1} |i\rangle \langle i | \otimes | j \rangle \langle j | $. The EW $W^{(+)}$ can detect entangled states in the one-parameter state family (isotropic states)  $\rho_{\alpha} = (1-\alpha)|\phi^{+}\rangle \langle \phi^{+} |+ \alpha \mathrm{I_2\otimes I_2} /4$ with the Bell state $|\phi^{+} \rangle =(|00\rangle + |11\rangle) / \sqrt{2}$ for $\alpha< 3/5$ (its know that for $\alpha<\frac{2}{3}$ the states are entangled). The mirrored one, $W^{(-)}$, detects entangled states in the class of (isotropic) states $\rho_{\beta} = (1-\beta)|\phi^{- }\rangle \langle \phi^{ - } |+ \beta \mathrm{I\otimes I} /4$ with the Bell state $|\phi^{-} \rangle =(|00\rangle - |11\rangle) / \sqrt{2}$ for $\beta < 1/3$. Note also that there is no EW that can detect entangled states $\rho_{\alpha}$ and $\rho_{\beta}$ at the same time. This is visualized in Fig.\ref{fig:simplex}. The (green) hyperplane corresponds to the witness $W^{(+)}$ and the (orange) hyperplane to the witness $W^{(-)}$. Both witnesses cover the full state space in this case. The upper and lower bounds are computed to be $U (W^{ (\pm) })=1/2$ with $p_{+} = 3/5$ and $p_{-}=7/5$ which give rise to the single observable as follows,
\begin{equation}  \label{eq:w+}
\widetilde{W} =  \frac{1}{10} \left( \begin{array}{cc|cc} 2 & 0 & 0 & -2 \\ 0 & 3  & 0 & 0 \\ \hline 0 & 0 & 3  & 0 \\ -2 & 0 & 0 & 2 \end{array} \right) ,
\end{equation}
for which the separability window is given by $[L (\ew),U (\ew)] =[3/20,7/20]$.}
\end{Example}

\begin{Example}[Two-Qutrits EWs] {\em The Choi EW $W^{(+)}$ obtained from the Choi map in $d=3$~\cite{ref:choimap} is given by
\begin{equation}
W^{(+)} = \frac{1}{6} \left( \sum_{i=0}^{2} [ 2| ii \rangle \langle ii |  + | i,i-1 \rangle \langle i, i-1 | ] - 3 \mathrm{P}_{+}  \right) \nonumber
\end{equation}
where $\mathrm{P}_+ = |\phi^+\rangle \langle \phi^+|$ with the Bell state $|\phi^+\rangle = (|00\rangle + |11\rangle + |22\rangle) / \sqrt{3}$. The Choi EW is known to be non-decomposable and can detect bound entangled states. One can find $U (W^{(+)})=2/9$ with $p_{+}=3/5$ and $p_{-} = 7/5$ so that the other one $W^{(-)}$ is given by
\begin{equation}
\,W^{(-)} = \frac{2}{9} \mathrm{I\otimes I}  - \widetilde{W}\;.
\end{equation}
Interestingly, while $W^{(+)}$ has a single  negative eigenvalue $-1/6$, its SPA-mirrored EW $W^{(-)}$ has two negative eigenvalues both $-1/9$. Note that both witnesses $W^{(\pm)}$ have the same upper bound $U(W^{(\pm)}) = 2/9$. Finally, the separability window $[L (\ew),U (\ew)]$ of the observable reads $[3/45,7/45]$. The p-SPA of the Choi EW is given by
\begin{equation}
\widetilde{W} = \frac 25\, W^{(+)} + \frac{3}{45} \,\mathrm{I\otimes I}\,
\end{equation}
which corresponds to a separable state as shown in Ref.~\cite{ref:korbicz}. Interestingly, $W^{(+)}$ is non-decomposable, whereas $W^{(-)}$ is decomposable.  }
\end{Example}

\begin{Example}[No Nontrivial Partner] {\em Here we show that not every EW $W^{(+)}$ has a nontrivial partner $W^{(-)}$. Consider $W^{(+)} = |\psi\rangle \langle \psi|^\Gamma$, where

\begin{equation}
  |\psi\rangle = \sum_i s_i\; |e_i\rangle \otimes |f_i\rangle ,
\end{equation}
denotes the Schmidt decomposition of $|\psi\rangle$. We assume $s_1 \geq s_2 \geq \ldots \geq s_d$. One finds $U(W^{(+)}) = \lambda_{\rm max} = s_1^2$, and hence

\begin{equation}
  W^{(-)} = \frac{1}{D s_1^2 -1} ( s_1^2\; \mathrm{I}_1 \otimes \mathrm{I}_2 - |\psi\rangle \langle \psi|^\Gamma) ,
\end{equation}
is a positive operator and hence cannot be be used as a witness.}
\end{Example}

\begin{Example}[Genuine Multipartite EWs]  {\em In Ref.~\cite{HMGH1} a general formalism to detect various types of multipartite entanglement was introduced which works for any number of particles and dimensions. It turned out to be efficient to guarantee the security in quantum secret sharing protocols~\cite{Hiesmayr_SecretSharing} or to prove experimentally the genuine multipartite entanglement of four physical photons entangled in their orbital angular momentum degrees of freedom~\cite{Hiesmayr_OAM} or of two photons entangled in polarisation and orbital angular momentum degrees of freedom~\cite{Hiesmayr_GHZ}. Let us here consider three qubits. For instance, the function
\begin{equation}
Q_{\rm GHZ}(\rho):=2 (\sqrt{\rho_{22}\rho_{77}}+\sqrt{\rho_{33}\rho_{66}}+\sqrt{\rho_{44}\rho_{55}}-|\rho_{18}|)\nonumber
\end{equation}
(with $\rho_{ij}$ being the elements of the density matrix $\rho$) is greater equal zero for all fully separable as well as for all biseparable states. Thus a violation of $0\leq Q_{\rm GHZ}(\rho)$ proves that the state $\rho$ is genuinely multipartite entangled. Moreover, the function is maximized for the Greenberger-Horne-Zeilinger state, e.g. $|GHZ\rangle=\frac{1}{\sqrt{2}}(|000\rangle+|111\rangle)$. The factor $2$ is chosen to set $Q_{\rm GHZ}(|GHZ\rangle\langle GHZ|)=\pm 1$. Now let us apply our procedure to obtain a new bound. Different to what we considered so far is that this inequality correspond to a nonlinear witness, but of course this witness can be linearised by using (i) $\sqrt{\rho_{ii}\rho_{jj}}\leq\frac{\rho_{ii}+\rho_{jj}}{2}$ and (ii) assuming $\rho_{18}$ to be purely real or imaginary. This gives $Q_{GHZ}^{lin}= 1-\rho_{11}-\rho_{88}+2Re\{\rho_{18}\}$ and obviously still the same maxima $\pm1$ for $|GHZ\rangle$. But considering the optimization over all pure fully separable states $\sigma_{sep}$ results in
\begin{eqnarray}
0 \leq &Q_{GHZ}(\sigma_{sep})&\leq\frac{1}{2} ,\nonumber\\
0 \leq &Q_{GHZ}^{lin}(\sigma_{sep})&\leq\;1 .
\end{eqnarray}
Consequently, the upper bound (mirrored witness)} looses its predictive power.

Another physically distinct genuine multipartite entangled state is the $W$-state or Dicke-state, e.g. in the computational basis given by $|Dicke\rangle=\frac{1}{\sqrt{3}}\{|001\rangle+|010\rangle+|100\rangle\}$ and the function~\cite{HMGH1}
\bea
Q_{Dicke}(\rho)&=&2(\frac{\rho_{22}+\rho_{33}+\rho_{55}}{2}
+\sqrt{\rho_{11}\rho_{44}}+\sqrt{\rho_{11}\rho_{66}}\nonumber\\
&&
+\sqrt{\rho_{11}\rho_{77}}-|\rho_{23}|-|\rho_{25}|-|\rho_{35}|)\;,\nonumber
\eea
which gives $-1$ for the Dicke state $|Dicke\rangle$ in the computational basis. However, note that though $|GHZ\rangle$ is also a genuine multipartite entangled state, but the minimum over all local unitaries leads only to $\min Q_{Dicke}(|GHZ\rangle)=-\frac{3}{4})$. Such as an optimisation over all local unitaries of the GHZ-witness function $Q_{GHZ}(|Dicke\rangle)=-0.628$ which does not equal the maximal value $-1$.

Now optimizing over all pure separable states of $Q_{Dicke}$ leads to
\bea
0&\leq& Q_{Dicke}(\sigma_{sep})\leq 1\;.
\eea
Indeed the upper bound is again a non-trivial one since $\max Q_{Dicke}(\rho_{ent})=1.5$ (actually the $\rho_{ent}$ equals $|GHZ\rangle$). Thus this expresses a certain duality between the two possible genuinely multipartite entangled states. Of course, one can also consider the optimization over all fully separable and bi-separable states, then, however, both criteria $Q_{GHZ/Dicke}$ provide no longer nontrivial upper bounds, they are optimal in this sense.
\end{Example}


\section{Extension to nonlinear witnesses}

EWs can be found as linear hyperplanes that distinguish entangled states from the set of separable states on the space of Hermitian operators. Since the set of separable states is convex, one may improve EWs by introducing nonlinearities such that the convex set is better characterized and a larger set of entangled states can be detected. For instance, it has been found that uncertainty relations in terms of variances offer such a possibility \cite{ref:lur, ref:var}.


A systematic method of constructing nonlinear EWs has been introduced \cite{ref:lut}. Suppose that some entangled state $\rho$ is detected by  a positive map $\Lambda$, that is, $[{\rm id} \otimes\Lambda](\rho)$ possesses an eigenvector $|\varphi\rangle \in \H_1 \otimes \H_2$ corresponding to a negative eigenvalue. Then the following formula

\bea
W = (\i \otimes \Lambda^{*})[ |\varphi\rangle \langle \varphi |] \label{eq:ewv}
\eea
where $\Lambda^{*}$ denotes a dual map, defines an EW detecting $\rho$, i.e. $\tr[W \rho]<0$. A nonlinear functional is then constructed by subtracting polynomials from an EW. To be specific, let us consider subtracting a single nonlinear polynomial as follows \cite{ref:lut},
\bea \label{N2}
{\cal F}^{(1)} (\rho) = \tr[ W \rho] - \frac{1}{s^2(\chi)} \tr [ \widetilde{X} \rho]\tr[ \widetilde{X}^\dagger \rho],
\eea
where one can choose $\widetilde{X} =  ({\rm id} \otimes \Lambda^*)( |\varphi\rangle \langle \chi|)$ for some state $|\chi \rangle \in \H_1 \otimes \H_2$ and $s(\chi)$ is the largest Schmidt coefficient of $|\chi \rangle$. The parameters are chosen such that $\mathcal{F}^{(1)}[\sigma]\geq 0$ for all $\sigma\in S_{\mathrm{sep}}$. It is clear that a witness $W$ does not detect a larger set of entangled states than its nonlinear functional ${\cal F}$ since $\mathcal{F}[\rho] \leq \tr[W\rho]$.

In what follows, we revisit mirrored EWs in Example 1 to demonstrate the improvements and illustrate construction of a nonlinear functional from an observable $A$ that links mirrored EWs. With respect to an observable $A$, the improvement can be made by both adding and subtracting polynomials.
\\


{\bf Example 5} (Nonlinear mirrored EWs). We recall that entangled states $|\phi^{\pm} \rangle = (|00\rangle\pm |11\rangle )/\sqrt{2}$ are detected by mirrored EWs $W^{(\pm)}$ in Example 1,
\bea
\tr[W^{( \pm)} |\phi^{\pm} \rangle \langle \phi^{\pm}|] = -1/4 \mp 1/8. \nonumber
\eea
From the EWs, positive maps $\Lambda^*_{(\pm)} $ are denoted as follows, see also Eq. (\ref{eq:ewv})
\bea
W^{(\pm)} = [{\rm id} \otimes \Lambda^*_{(\pm)}]( |\phi^{\pm}\rangle \langle \phi^{\pm}|).\nonumber
\eea
According to Eq. (\ref{N2}), a nonlinear polynomial can be constructed as $\widetilde{X}^{(\pm)} =  ({\rm id} \otimes \Lambda_{(\pm)}^*)( |\phi^{ \pm }\rangle \langle \phi^{\mp}|)$, where we have chosen $|\chi\rangle = | \phi^{\mp}\rangle$,
\bea
\widetilde{X}^{(+)} = \frac 18 \left( \begin{array}{cc|cc} 1 & 0 & 0 & -4 \\ 0 & 3 & 0 & 0 \\ \hline 0 & 0 & -3 & 0 \\ 4 & 0 & 0 & -1 \end{array} \right),
  \widetilde{X}^{(-)} = \frac{1}{8} \left( \begin{array}{cc|cc} 3 & 0 & 0 & -4 \\ 0 & 1 & 0 & 0 \\ \hline 0 & 0 & -1 & 0 \\ 4 & 0 & 0 & -3 \end{array} \right)\ .\nonumber
\eea
Since $|\chi\rangle = | \phi^{\mp}\rangle$, we have $s^2(\chi) =1/2$. Then, a pair of nonlinear EWs are obtained from mirrored EWs,
\bea
{\cal F}_{(\pm)}^{(1)} (\rho) = \tr [W^{(\pm)} \rho] - 2 \tr [ \widetilde{X}^{(\pm)} \rho] ~\tr [ \widetilde{X}^{(\pm)})^\dagger \rho], \label{N3}
\eea
which is non-negative for all separable states \cite{ref:lut}.

For instance, we consider a state $|\mu \rangle  = a|00\rangle + b|11\rangle$ with $a,b\in \mathbbm{R}$ and $a^2+b^2=1$:
\bea
&& \tr[W^{(+)} |\mu\rangle \langle \mu |] = \frac{1}{8} (a^2 +b^2 -8ab), ~\mathrm{and}  \nonumber \\
&& \tr[W^{(-)} |\mu\rangle \langle \mu |] = \frac{1}{8} (3 a^2 +3 b^2 + 8ab). \nonumber
\eea
The state $|\mu\rangle$ with $ab=1/8$ is entangled but not detected by $W^{(+)}$ since $\tr[W^{(+)} |\mu\rangle \langle \mu |] =0$. It can be detected by the nonlinear witness in Eq. (\ref{N3}): we have that ${\cal F}_{(+)}^{(1)} ( |\mu\rangle \langle \mu| ) = -1/64 <0$. For $ab = -3/8$, we have $\tr[W^{(-)} |\mu\rangle \langle \mu |] =0$ but ${\cal F}_{(-)}^{(1)} ( |\mu\rangle \langle \mu| ) = -9/8 <0$. Thus, the improvement by nonlinear polynomials in Eq. (\ref{N3}) is shown.

We recall that the mirrored witnesses $W^{(\pm)}$ characterize the lower and upper bounds of an observable $A$ for separable states, see Eqs. (\ref{eq:main}), (\ref{eq:w-a}), and (\ref{eq:w+a}).  It is straightforward to find a nonlinear functional with the observable $A$ that links mirrored EWs $W^{(\pm)}$ from Eq. (\ref{N3}): we write by $\mathcal{G}$ a nonlinear functional from an observable $A$ via nonlinear EWs in Eq. (\ref{N3})
\bea
\mathcal{G}_{\pm} [\rho] = \tr[A \rho] \pm 2 {n_{\mp}}    \tr [ \widetilde{X}^{(\mp )} \rho] ~\tr [ \widetilde{X}^{( \mp)})^\dagger \rho] \label{eq:nlA}
\eea
where $n_{\pm}$ are from Eq. (\ref{eq:const}). Note that both addition and subtraction of nonlinear polynomials are essential to improve an observable $A$. It holds that for $\sigma\in S_{\mathrm{sep}}$,
\bea
\mathcal{F}_{(+)}^{(1)} [\sigma] \geq 0 & \iff & L(A) \leq \mathcal{G}_{-} [\sigma] \nonumber \\
\mathcal{F}_{(-)}^{(1)} [\sigma] \geq 0 & \iff &  \mathcal{G}_{+} [\sigma] \leq U(A). \label{eq:anlu}
\eea
This shows how to improve an observable $A$ with nonlinear polynomials. For the lower bound $L(A)$ for separable states, which is characterized by an EW $W^{(+)}$, the improvement is made by subtracting a nonlinear polynomial, i.e., $\mathcal{G}_{-}$ in Eq. (\ref{eq:nlA}). Note that this is similar to the way that a standard EW is improve by nonlinear polynomials, see Eq. (\ref{N3}). For the upper bound $U(A)$, however, the improvement is obtained by adding a nonlinear polynomial, i.e., $\mathcal{G}_{+}$ in Eq. (\ref{eq:nlA}). \\

The improvement of an observable $A$ can be rephrased on the level of standard EWs. The separability window in Eq.~(\ref{eq:main}) can be rewritten for all $\sigma \in S_{\mathrm{sep}}$ by
\bea
0\leq \tr[W\sigma] \leq U(W).  \label{eq:ub}
\eea
Note that from Eqs.~(\ref{eq:w-a}) and (\ref{eq:equi}), there exists a witness $W^{'}$ that characterizes the upper bound, i.e., for any state $\rho$ on $\H_1 \otimes \H_2$, it holds that
\bea
\tr[W\rho] \leq U(W) \iff 0\leq \tr[W^{'} \rho]\;. \nonumber
\eea
 Let us recall that the detection method by a lower bound $\tr[W\sigma] \geq 0$ for $\sigma\in S_{\mathrm{sep}}$ is improved by subtracting nonlinear polynomials. The improvement in Eq.~(\ref{eq:anlu}) means the following statements are equivalent:
\begin{itemize}
\item the detection method by an upper bound
\bea
\tr[W\sigma] \leq U(W)~\mathrm{for}~ \sigma\in S_{\mathrm{sep}} \nonumber
\eea
is improved by adding nonlinear polynomials, and
\item the detection method by a lower bound
\bea
\tr[W^{'}\sigma] \geq 0~\mathrm{ for}~ \sigma\in S_{\mathrm{sep}}\nonumber
\eea
is improved by subtracting nonlinear polynomials.
\end{itemize}
Thus, it is shown that nonlinear EWs can be thus generally constructed by both subtracting and adding nonlinear polynomials. 



\begin{figure}[t]
\includegraphics[width=3.5in,keepaspectratio]{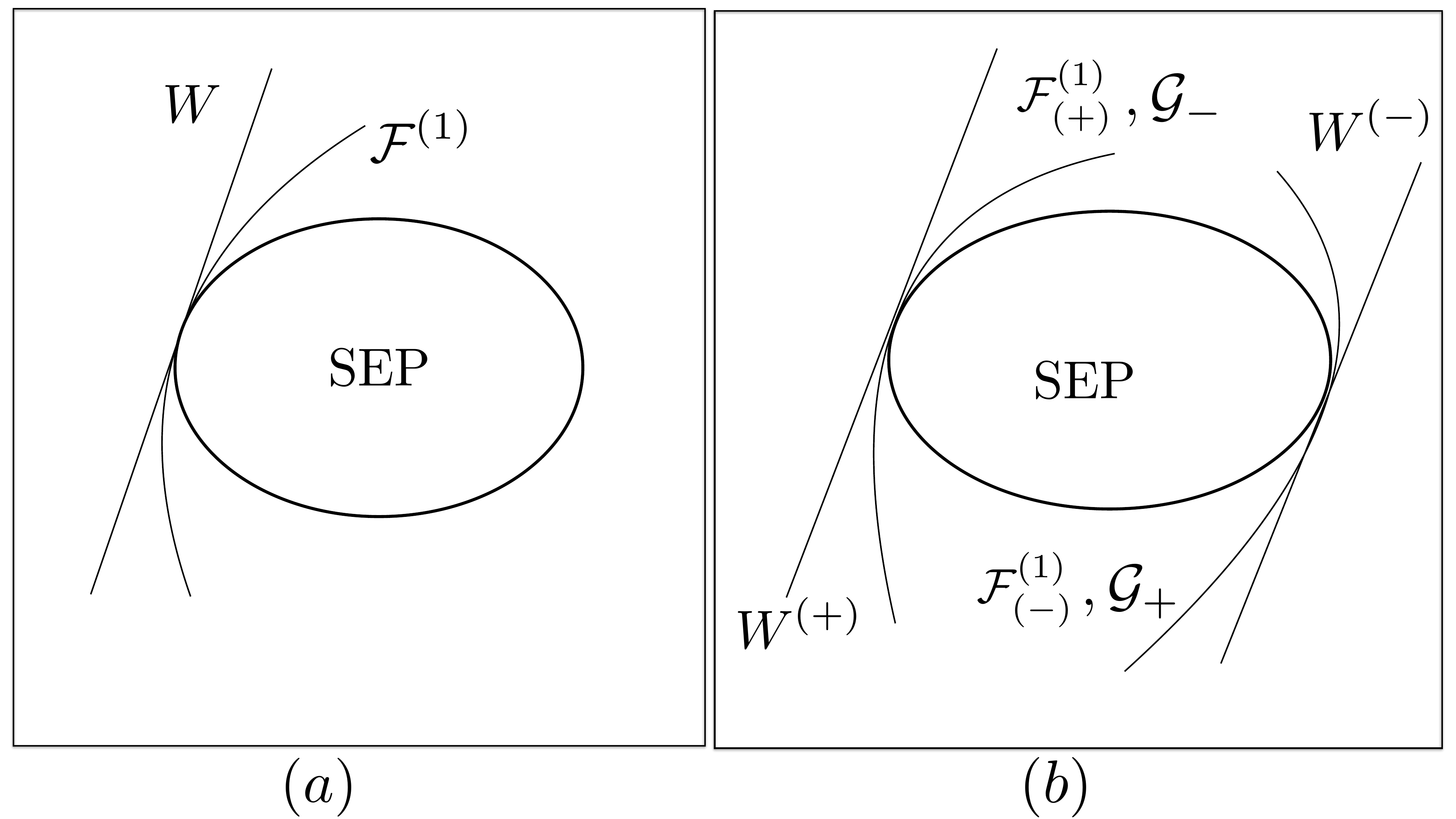}
\caption{ $(a)$ A nonlinear EW $\mathcal{F}^{(1)}$ can be obtained by subtracting a polynomial from a witness $W$, so that a larger set of entangled states can be detected \cite{ref:lut}, see Eq. (\ref{N3}). $(b)$ Mirrored EWs $W^{(-)}$ and $W^{(+)}$ that characterize upper and lower bounds of an observable for separable states, respectively, can also be improved, which are denoted by $\mathcal{F}_{(\pm)}^{(1)}$. For the observable, nonlinear EWs are introduced by both subtracting or adding a polynomial, see $\mathcal{G}_{\pm}$ in Eq. (\ref{eq:nlA}).
  }
\label{fig:3}
\end{figure}


\section{Summary and Outlook}

\begin{figure}[t]
\includegraphics[width=3.5in,keepaspectratio]{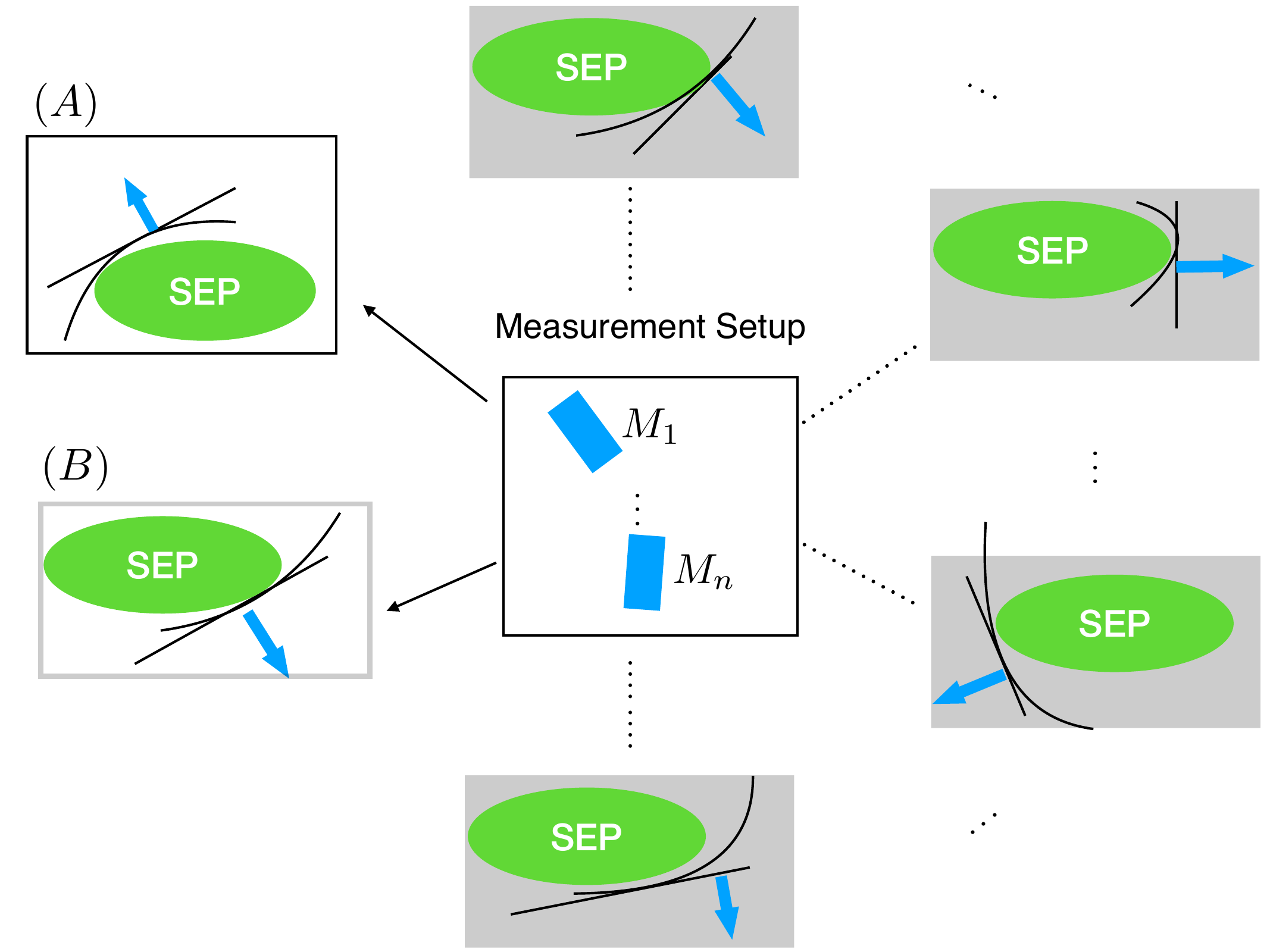}
\caption{An EW, Eq. (\ref{eq:lb}), corresponds to an observable that can be estimated in a measurement setup that contains a set of POVM elements. The box~$(A)$ shows a schematic structure of the entanglement detection with an EW and its nonlinearization. The present work shows that the box~$(B)$ exists: the measurement setup prepared for the EW in the box~$(A)$ can be used to construct the other EW in the box~$(B)$, i.e., its mirrored one. This EW detects other types of entangled states utilizing the same information from the experiment (center), i.e. no other experimental setup is needed. It remains open to disclose the gray boxes, i.e., the possibility of constructing more than two EWs from a single measurement setup (see also Sec. \ref{sec:sum}).}
\label{fig:4}
\end{figure}

We have presented the framework of mirrored EWs. It is shown that a single observable corresponds to a pair of EWs, called mirrored EWs, in the sense that entanglement detection with mirrored EWs can be performed by realizing a single observable in practice. Then, a single observable has both upper and lower bounds for separable states, where mirrored EWs  gives the characterization of the bounds, respectively. Thus, entangled states can be detected by violations of either of the bounds. It is shown that mirrored EWs can be constructed from bounds of an observable for separable states, as well as via structural physical approximations. Equivalently, mirrored EWs can be rephrased that an EW generally has a non-trivial upper bound for separable states, such that the upper bound is characterized by its pair. I.e., this shows that an EW can generate its paired one. Examples are presented to illustrate how the framework of mirrored EWs works in detail to detect different types of entangled states.

Finally, nonlinearties are introduced to improve mirrored EWs, i.e., mirrored nonlinear EWs. It is shown that both subtraction and addition of nonlinear polynomials can improve mirrored EWs: the subtraction improves standard EWs that detect entangled states from violations of the lower bound, and the addition does the other detection scheme with a non-trivial upper bound. We emphasize that our findings show, via the effort of mirrored EWs, the recorded experimental knowledge that has been used to detect entanglement from Eq.~(\ref{eq:lb}) can be utilized once again to see the violation of an upper bound in Eq.~(\ref{eq:ub}), without acquiring additional experiments. A larger set of entangled states than the one from the very definition of a witness can be detected.

The structure of SPA behind mirrored EWs can be generalized by replacing the identity operator with some other separable operators. That is, the operator $\mathrm{I_1\otimes I_2}/D$ in a  positive and negative SPAs in Eqs. (\ref{p}) and (\ref{q}) can be replaced with a full-rank, non-negative and unit-trace operator $X \in S_{\rm sep}({\cal H})$. Note that for a full-rank, non-negative and unit-trace operator $X$, there always exist  such $p_X$ and $q_X$  \cite{ref:remik}. In future investigations, it would be interesting to extend mirrored EWs by generalizing SPAs and to explore further structures of entangled states.

EWs are a useful tool not only to investigate theoretical and experimental aspects of entangled states but also to certify properties having a convex structure, such as fidelities of states or quantum operations, see e.g. \cite{ref:r1, ref:r2}. Our results, that the EWs can be indeed {\it mirrored,} pave a new avenue in the certification of entanglement and other quantum properties. In future investigations, it would be interesting to find the theoretical characterizations of mirrored EWs in the view of (non-)decomposability, extremality of EWs, etc. \cite{ref:review3, Optimal}. Although EWs apply single-copy level measurements that are feasible with current technologies, it is worth to extend the framework of mirrored EWs to multi-copy scenarios, e.g., with more assumptions on state preparation \cite{ref:enk} and with more advanced technologies, see e.g. \cite{ref:a1, ref:a2, ref:a3}. From a practical point of view it is also worth to consider the statistical errors in an implementation of nonlinear and mirrored EWs, e.g., the implementation should be constructed such that the ranges of statistical errors of the paired EWs do not overlap each other within the separability window.

%

\section{Acknowledgement} J.B. is supported by National Research Foundation of Korea (2019M3E4A1080001), the KIST Institutional Program (2E29580-19-148), an Institute of Information and Communications Technology Promotion (IITP) grant funded by the Korean government (MSIP), (Grant No. 2019-0-00831, EQGIS) and ITRC Program (IITP-2019-2018-0-01402) and BK21PLUS.  D.C. was supported by the Polish National Science Centre project 2015/19/B/ST1/03095. B.C.H. acknowledges gratefully the Austrian Science Fund (FWF-P26783).

%

%
%
%


\end{document}